\documentclass[letter,english,superscriptaddress,reprint,prl,noshowkeyws,showpacs, nobibnotes]{revtex4-1}

\usepackage[T1]{fontenc}
\usepackage[latin9]{inputenc}
\usepackage{amsmath}
\usepackage{graphicx}
\usepackage{amssymb}
\usepackage{esint}
\usepackage{setspace}
\usepackage{babel}
\usepackage[normalem]{ulem}
\usepackage{amsmath,amstext,amsbsy,amsopn,amsthm,amscd,amsfonts,amssymb}
\usepackage{bm}
\usepackage{epsfig}
\usepackage{graphpap}
\usepackage{mathrsfs}
\usepackage{setspace}
\usepackage{subfigure}

\theoremstyle{definition}

\theoremstyle{remark}

\newcommand{\REMOVE}[1]{}

\newcommand{\begeq}{\begin{equation}}
\newcommand{\eneq}{\end{equation}}
\newcommand{\beq}{\begin{equation}}
\newcommand{\eeq}{\end{equation}}
\newcommand{\beqa}{\begin{eqnarray}}
\newcommand{\eeqa}{\end{eqnarray}}

\newcommand{\um}{{\mu{m}}}
\newcommand{\fig}[1]{Fig.~\ref{#1}}

\newcommand{\BSCCO}{${\rm Bi_2Sr_2CaCu_2O_{8+\delta}}~$}

\newcommand{\Tc}{${\rm T_c}$~}

\newcommand{\Hab}{H_{ab}}

\begin{document}
\title{Lamellar mesophase nucleated by Josephson vortices at the melting of the
vortex lattice in \BSCCO}
 \author{Y. Segev}
\homepage{http://www.weizmann.ac.il/condmat/superc/}
\affiliation{Department of Condensed Matter Physics,
Weizmann Institute of Science, Rehovot 76100, Israel}


\author{Y. Myasoedov}
\affiliation{Department of Condensed Matter Physics,
Weizmann Institute of Science, Rehovot 76100, Israel}

\author{E. Zeldov}
\affiliation{Department of Condensed Matter Physics,
Weizmann Institute of Science, Rehovot 76100, Israel}

\author{T. Tamegai}
\affiliation{Department of Applied Physics, The University of Tokyo, Hongo, Bunkyo-ku, Tokyo 113-8656, Japan}

\author{G. P. Mikitik}
\affiliation{B. Verkin Institute for Low Temperature Physics \& Engineering, National Ukrainian Academy of Sciences, Kharkov 61103, Ukraine}

\author{E. H. Brandt}
\thanks{Deceased}
\affiliation{Max-Planck-Institut f\"{u}r Metallforschung, D-70506 Stuttgart, Germany}

\begin{abstract}
The local effect of the Josephson vortices on the vortex lattice melting process in \BSCCO crystals in the presence of an in-plane field $H_{ab}$ is studied by differential magneto-optical imaging. The melting process is found to commence along the Josephson vortex stacks, forming a mesomorphic phase of periodic liquid and solid lamellas, the direction and spacing of which are controlled by $H_{ab}$. The reduction of the local melting field $H_m$ along the Josephson vortex stacks is more than an order of magnitude larger than the reduction of the average bulk $H_m$ by $H_{ab}$.
\end{abstract}
 \maketitle 

\maketitle
\label{meltingJV}
\label{meltingJV-introduction}

The melting of the vortex lattice in type II
superconductors (SCs) \cite{Blatter-1994,*Mikitik-2001a} is
an exemplary phase transition that allows direct imaging of
the melting process with control over the inter-particle
interaction, degree of quenched disorder, and the
anisotropy of the host crystal. This phase transition was
extensively studied in \BSCCO \;(BSCCO), which is a layered
high-\Tc SC with very high anisotropy
$\gamma=\lambda_c/\lambda_{ab}\approx 500$, where
$\lambda_c$ and $\lambda_{ab}$ are out of plane and
in-plane penetration depths. In such layered SCs the
energetic cost of penetration of an in-plane field $\Hab$
is so low that the SC becomes almost ``transparent'' to
$\Hab$. As a result, the macroscopic magnetic response of
the SC, as well as the vortex lattice melting transition
$H_m(T)$, are mainly governed by the c-axis field $H_c$.
Experimental and theoretical studies have accordingly shown
that in order to appreciably change the melting field $H_m$
in such layered systems, $\Hab$ has to be two to three
orders of magnitude larger than $H_c$. In BSCCO, in
particular, $H_m$ was found to decrease linearly with
$\Hab$, $H_m=H_m^0-\alpha\Hab$ with $\alpha\sim0.01$,  for
not too high $\Hab$
\cite{Schmidt-1997,*Ooi-1999,*Mirkovic-2001,Koshelev-1999}.

\begin{figure}[b]
            \centering
            \includegraphics[width =0.48 \textwidth%
            ,bb=  71 41 590 365, clip=,
            ]{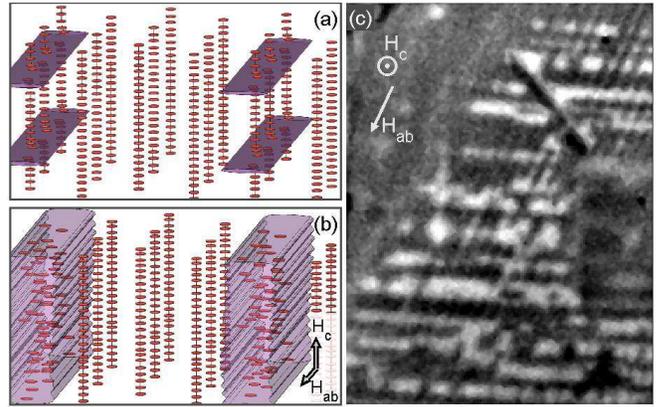}
            \caption{(a) Solid crossing lattices state of pancake (red) and
            Josephson (purple) vortices. PVs residing along the JVs
            form chain structure whereas PV stacks in-between the JV
            planes form an Abrikosov lattice. For clarity, the spacing and dimensions of the JVs are shown
            not to scale. (b) Schematics of lamellar mesophase
            in which a solid PV lattice coexists with a PV liquid
            along the JV planes. In the liquid lamellas both the PVs and
            the JVs may be fully dissociated forming `gas' lamellas. (c) A field modulated DMO image
            showing preferential nucleation of melting along the JVs (bright diagonal lines) at
            $H_c=15$~Oe, $\delta H=0.5$~Oe,
            $\Hab=30$~Oe, $\varphi=30^\circ$, and
            $T=86.5$~K. The characteristic horizontal features are due to sample inhomogeneities.
            The image is 0.3 mm wide.
            }
            \label{schematics}
\end{figure}

In contrast to its very weak average effect, $\Hab$ has a
major impact on the \emph{local} vortex lattice structure.
While $H_c$ results in an hexagonal lattice of
two-dimensional pancake vortices (PVs) \cite{Clem-1991},
$\Hab$ gives rise to an oblate lattice of Josephson vortex
(JV) stacks which may coexist with the PV lattice forming a
\emph{crossing lattices} state
\cite{Koshelev-1999,Bending-2005} with an attractive
interaction between the PVs and JV stacks. As a result, the
hexagonal PV lattice is replaced by one of the following
configurations, depending on the microscopic parameters and
the densities of PVs and JVs
\cite{Bulaevskii-1992,Koshelev-1999,Koshelev-2005}. At low
$H_c$, the \emph{chain state} is obtained in which the PV
stacks reside only along the 2D planes formed by the JV
stacks (parallel to $H_c$ and $\Hab$), which we refer to as
JV planes (JVPs). At high $H_c$, the JVPs cause only a weak
perturbation, resulting in a smooth modulation of the
vortex density in the 3D lattice in the direction
perpendicular to the JVPs --- \emph{modulated Abrikosov
lattice}. The most interesting regime, however, occurs at
intermediate values of $H_c$, in which vortex chains
residing on the JVPs coexist with a 3D Abrikosov lattice
that occupies the ``bulk'' between the JVPs forming a
\emph{mixed chain and lattice state} as observed by
decorations
\cite{Bolle-1991,*Grigorieva-1995,*Tokunaga-2003},
magneto-optics
\cite{Vlasko-Vlasov-2002,*Tokunaga-2002,Segev-2011},
scanning Hall probe
\cite{Grigorenko-2001,*Grigorenko-2002}, and Lorentz
microscopy \cite{Matsuda-2001}. In this mixed regime, Fig.
1a, a \emph{inhomogeneous} binary system is formed,
consisting of two qualitatively different subsystems with
different dimensionalities, local effective anisotropies,
and elastic moduli. So far, the melting transition in
presence of $\Hab$ was treated only in an effective
``homogeneous medium'' approach assuming a single bulk $H_m(T)$
\cite{Koshelev-1999}. However, since the two vortex
subsystems are significantly different, an emerging
fundamental question is whether such a binary system,
instead of having a single transition, may display two
separate transitions --- one for each of the subsystems.
This would imply a phase separation on the scale
of a single vortex lattice unit cell, resulting in an intermediate periodic
solid-liquid mesophase with a lamellar structure, see Fig.
1b.

In this work we address this question experimentally. We find that
JVs significantly alter the PV melting landscape resulting in two
distinctive melting processes. The JVPs cause a \emph{local} reduction in
$H_m(T)$ which is an order of magnitude larger than the average
effect of $\Hab$. As a result, a periodic solid-liquid structure is
formed similar to lamellar mesophases in liquid crystals, silicates,
and
various organic compositions 
\cite{Monnier-1993,*Gabriel-2001}. The characteristic
feature of these molecular mesophases is their
heterogeneous chemical composition. In contrast, the vortex
lattice is a unique example of an initially homogeneous
system that can be readily driven into a lamellar
solid-liquid structure in which the direction, spacing, and
thickness of the two types of lamellas can be controlled by
tilting the applied magnetic field.

\begin{figure}[t]
            \centering
            \includegraphics[width =0.45 \textwidth,
            bb=  21 43 585 469, clip=,]{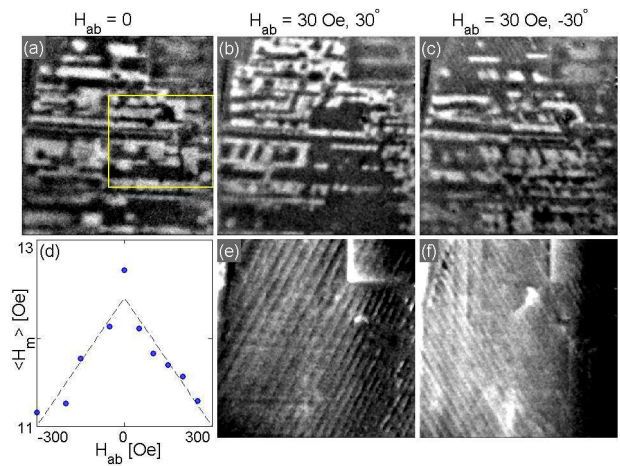}
            \caption{Field modulated DMO images ($\delta H=0.5$~Oe) of the melting transition at $T=84$ K, $H_c=24$ Oe, $\Hab=0$ (a), $T=85$ K, $H_c=20$ Oe, $\Hab=30$ Oe, $\varphi=30^\circ$ (b), and $T=85$ K, $H_c=20.5$ Oe, $\Hab=30$ Oe, $\varphi=-30^\circ$  (c). (e) and (f) DMO images of PV chains at $H_c=3$ Oe and $H_c=7$ Oe at the same $\Hab$ and $T$ as (b) and (c), respectively. The images are 0.5 mm wide. (d) The average melting field $H_m$ vs. $\Hab$ at $T=87$ K ($\circ$) and a linear fit $H_m= 12.4 {\rm{Oe}}-4\times 10^{-3}|\Hab|$.}
            \label{fig1}
\end{figure}

Magneto-optics was previously used for imaging PV-decorated
JVs
\cite{Vlasko-Vlasov-2002,*Tokunaga-2002,Yasugaki-2002,Segev-2011}
and vortex lattice melting
\cite{Soibel-2000,Yasugaki-2002,Soibel-2001,*Yasugaki-2003,Banerjee-2004,
Goldberg-2009}. Here we use differential magneto-optics
(DMO) for microscopic imaging of the melting process in the
presence of JVs. The data were obtained on an optimally
doped BSCCO crystal of $2750\times740\times30~\mu\rm{m}^3$
with $T_c\approx 91$~K and anisotropy $\gamma\approx 500$.
Qualitatively similar results were obtained on several
other optimally doped samples. In field-modulated (or
temperature-modulated) DMO the difference of two images of
the local induction $B(x,y)$ taken at $T_0,H_0\pm\delta H$
(or $T_0\pm\delta T,H_0$) is acquired and averaged $\sim10$
times. The resulting DMO images visualize the \emph{local}
$dB/dH(x,y)$ (or $dB/dT(x,y)$). Since the first-order
melting $H_m(T)$ is accompanied by a step-like
discontinuity in $B$ \cite{Zeldov-1995} and hence by a peak
in $dB/dH$ and $dB/dT$ \cite{Morozov-1996a}, regions in the
sample that are driven through the phase transition by the
modulation $2\delta H$ (or $2\delta T$) display in DMO
images significantly enhanced $dB/dH$ (or $dB/dT$) relative
to the neighboring regions that remain in the same phase
\cite{Soibel-2000}. Figure \ref{fig1}a shows a typical
field-modulated DMO image of the melting process in BSCCO.
The bright broad patches are the regions that undergo
melting at $T=84$ K and $H_c=24$ Oe within field modulation
of 1 Oe. The characteristic long (horizontal) lines are
caused by very small compositional inhomogeneities
originating from the floating-zone growth process of BSCCO
crystals giving rise to small variations in the local
$H_m(T)$ as described previously
\cite{Soibel-2001,Yasugaki-2003}. As shown below, this
quenched-disorder-induced landscape is dramatically altered
in the presence of $\Hab$.

We first evaluate the \emph{mean} effect of $\Hab$ by calculating
the average melting field $\left< H_m \right>$ from a
series of DMO images vs. $H_c$ at various $\Hab$. Figure
\ref{fig1}d shows the very weak monotonic decrease of
$\left< H_m \right>$ with $\Hab$ that can be described by
$\left< H_m \right> \simeq \left< H_m \right>_0 -
4\times10^{-3}|\Hab|$, consistent with previous reports
\cite{Schmidt-1997,*Ooi-1999, Koshelev-1999}. The \emph{local} effect of the JVs on the melting
landscape is demonstrated in Figs. \ref{fig1}b and \ref{fig1}c, which show the
melting transition for $\Hab = 30$ Oe, applied at in-plane
angles $\varphi$ of $30^\circ$ and $-30^\circ$ relative to
the vertical axis of the images. The melting patterns in the presence of $\Hab$
are markedly different from \fig{fig1}a having much finer structure and
pronounced diagonal orientation along the direction of
$\Hab$. For comparison, Figs. \ref{fig1}e and \ref{fig1}f
show DMO images at the same conditions but at lower values
of $H_c$ at which the solid chain state is visible as
diagonal lines crossing the sample. The comparison of the
melting and chain images clearly demonstrates that the JVs
have a strong \emph{local} effect on the melting transition
rather than just a weak uniform suppression of $H_m$. They
locally alter the mechanism and the effective
dimensionality of the melting nucleation and propagation
processes inducing periodic structure that has the same
spacing and orientation as the JV stacks. Figure
\ref{schematics}c shows an expanded view of the diagonal
melting patterns at a higher temperature
and the corresponding full movie of the melting process is
available online \cite{AuxWebPage}.

\begin{figure}[t]
            \centering
            \includegraphics[width =0.49 \textwidth,
            bb=  24 26 584 408,clip=,]{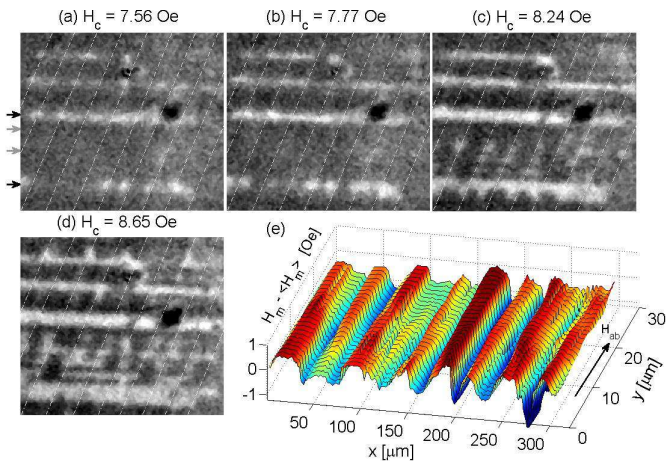}
            \caption{(a--d) Sequence of temperature modulated DMO images ($\delta T=0.2$ K)
            showing the progression of the melting transition with $H_c$ at $T=87$~K and $\Hab=5$ Oe, $\varphi=30^\circ$ in the marked region in \fig{fig1}a. The dashed lines are guide
            to the eye for the location of JVPs. The images are $270\;\mu m$ wide. (e) The measured variation of the local melting field $H_m$ across the JVPs showing a modulation of $0.7\pm 0.2$ Oe.}
            \label{fig3}
\end{figure}

The interplay between the sample growth inhomogeneities and
the JV lattice is demonstrated in detail in \fig{fig3} that
shows a magnified view (yellow rectangle in \fig{fig1}a) of
the melting transition by sweeping $H_c$ in the presence of
a widely spaced JV lattice, $\Hab=5$ Oe,
$\varphi=30^\circ$. The figure shows that melting nucleates
first at the crossings points between the JVPs and the
horizontal growth inhomogeneities (e.g. along the lower
black arrows in \fig{fig3}a). In \fig{fig3}b the melting
expands in-between the JVPs but only within the horizontal
defects. In the next step, \fig{fig3}c, the melting
propagates into the rather uniform wider regions in-between
the horizontal defect lines (e.g. the region between the
two black arrows in \fig{fig3}a). Interestingly, within
these homogeneous regions the melting starts along the JVP
forming periodic structure of liquid lamellas with a period
of about 35 $\um$. The initial apparent thickness of the
liquid lamellas is very small $\sim 4\;\um$, limited by the
spatial resolution of our system. With increasing $H_c$ the
liquid lamellas expand, \fig{fig3}d, followed by melting of
the rest of the lattice at still higher $H_c$.

In order to quantify the effect of the JV lattice on the
melting, we perform a very fine scan of $H_c$ with 50 mOe
steps and find the melting field $H_m(T,\Hab;x,y)$ at each
pixel, defined as the first $H_c$ at which the pixel
diffrential intensity reaches a threshold value. Figure
\ref{fig3}e shows the resulting melting landscape along a
narrow strip between the two gray arrows in \fig{fig3}a,
from which a smoothed, averaged, $\left<{H}_m\right>$ has
been subtracted for clarity. The modulation of $H_m$ by the
JVPs is clearly visible with a modulation amplitude of
$\Delta H_m=0.7\pm0.2$~Oe. Note that in order to achieve
such $\Delta H_m$ shift within the homogeneous medium
approach, one requires $\Hab\approx200$ Oe (see e.g.
\fig{fig1}d), whereas \fig{fig3}e shows such a reduction,
locally, with $\Hab=5$~Oe. Thus the \emph{local} modulation
of $H_m$ by the JVs is more than an order of magnitude
larger than their average effect. A similar analysis was
performed on the local behavior of $T_m(H_c,\Hab;x,y)$
using a fine temperature scan, where it was found that the
JVs induce a modulation $\Delta T_m\simeq0.1$ K. The two
results are consistent taking into account the slope of the
melting line $\partial H_m/\partial T_m \approx 5 \; {\rm
Oe/K}$ in optimally doped BSCCO close to $T_c$
\cite{Zeldov-1995,Beidenkopf-2007}. Another interesting
observation in \fig{fig3}e is the shape of the $H_m$
modulation, which often shows a sharp V-shaped dip at the
JVPs and a rather flat maximum between the dips. This
implies that the suppression of $H_m$ by the JVPs is very
local resulting in formation of liquid lamellas (see Fig.
1b) that can be as thin as a `monolayer' of PV stacks which
is about 1--2\;$\um$ within our experimental conditions and
is below our spacial resolution.

Our results can be summarized as following: (i) JVPs
locally suppress $H_m(T)$ of the PV lattice forming a
mesomorphic phase of parallel solid-liquid lamellas with a
direction and periodicity that are controlled by $\Hab$.
(ii)~Quenched disorder due to variations in crystal growth
conditions introduce large scale variations in local
$H_m(T)$ while JVs cause a controllable secondary melting
corrugation on a fine scale. (iii) In more disordered
regions, vortex liquid droplets are formed at intersection
points between the JVs and the growth disorder planes. (iv)
The liquid lamellas on the JVPs appear to maintain their
alignment even if separated by macroscopic liquid regions.

The melting transition occurs when the induction $B_c$
reaches $B_m(T)$. Since the density of PVs and hence $B_c$
along the JVPs is higher than in-between them, one may
argue that the observed melting along the JVs is a result
of the enhancement of the local $B_c$ by $\sim0.7$~G,
rather than JV-induced suppression of the local $B_m$. We
find that the enhancement of $B_c$ on the JV stacks is
large only close to the penetration field \cite{Segev-2011}
as seen in Figs. \ref{fig1}e and \ref{fig1}f. Upon
increasing $H_c$, the DMO visibility of the JV stacks
decreases monotonically and vanishes well below $H_m$,
whereas a 0.7 G spatial modulation in $B_c$ should have
been well within our experimental resolution even in
non-differential images. We therefore, conclude that the
dominant mechanism of formation of the lamellar structure
is the modulation of the local $B_m$ by the JV lattice.

A priori, it is not obvious that the melting should
nucleate on the JV stacks. The stiff binding potential of
the JVP might suppress thermal fluctuations of the PV chain
and hence shift $H_m$ to higher fields, similar to the
$H_m$ enhancement found in BSCCO in the presence of a low
concentration of columnar defects \cite{Banerjee-2004}. On
the other hand, Lorentz microscopy has shown that in very
thin BSCCO, the PV vibrations along the chains are
significantly enhanced due to incommensurability between
the chain and the adjacent lattice \cite{Matsuda-2001},
which can explain the experimental observation of a
reduction of $H_m(T)$ on the chains. The magnitude of the
local reduction of $H_m$ on the JVPs  can be roughly
compared with the global measurements as following. In
global measurements the separate melting transitions of the
chains and the lattices will be observable as single
broadened transition. Assuming that the JVs suppress the
melting only within their core region $\lambda_{J0}=\gamma
s \approx 0.75\;\um$ ($s$ is the interlayer spacing), we
can evaluate the global $\Delta H_m$ as a volume weighted
average of the local transitions. In \fig{fig3}, at
$\Hab=5$ Oe, the local melting suppression on the JVPs is
$\Delta H_m=0.7$ Oe, and the average distance between the
JVPs is $a_J=36 \mu$m, resulting in a weighted global
$\Delta H_m=(\lambda_{J0}/a_J)\times0.7~\rm{Oe}
=1.5\times10^{-2}$ Oe. Assuming a linear $\Hab$ dependence
we obtain $\alpha=\Delta H_m/\Hab=3\times10^{-3}$, which is
in a general agreement with published results
\cite{Schmidt-1997,*Ooi-1999,*Mirkovic-2001,Koshelev-1999}
and \fig{fig1}d.

In our experimental range of $B_c\lesssim45$ G the
intervortex spacing is larger than the JV core,
$a_0>\lambda_{J0}$, so that each JVP is occupied by a
single row of PVs. This fact, combined with the observation
that the melting nucleates along the JVPs, suggests that at
the onset of melting liquid `vortex monolayer' lamellas are
formed along the JVPs that are `sandwiched' between thicker
lamellas of vortex solid. This may present a very
interesting case of interacting vortex lines in a
`1+1'-dimensional random potential or of a Luttinger liquid
\cite{Bolle-1999,Nattermann-2003}. The properties of such a
liquid monolayer of lines is unclear, in particular taking
into account the incommensurate periodic potential induced
by the two adjacent solids. Moreover, the first-order
melting in BSCCO is believed to be a `sublimation'
transition--- simultaneous melting and layer decoupling---
resulting in a PV `gas' with no c-axis correlations
\cite{Blatter-1996,*Fuchs-1997,*Horovitz-1998,*Shibauchi-1999,Koshelev-1999}.
This constitutes an even more intriguing situation in which
`gas' lamellas are caged between solid lamellas.
Remarkably, in such a mesostate the JVs may cease to exist
due to the random motion of the PVs along the `gas'
lamellas resulting in Josephson `sheets', where the
$\rm{CuO_2}$ planes are fully decoupled in the `gas'
monolayers and fully coupled in the solid regions. This
situation is schematically depicted in \fig{schematics}b.

Our finding of the heterogeneous behavior questions the
validity of existing approaches in which $\Hab$ is
considered as a source of a uniform $H_m(T)$ reduction in
an effective homogeneous medium \cite{Koshelev-1999}. In
particular, if one assumes that the suppression of the
local $H_m(T)$ is related to the local density of JVs or of
the crossing points between the JVs and the PV stacks, one
may expect $\Delta H_m\propto\sqrt{\Hab}$ along the JVPs
and $\Delta H_m=0$ in-between them. Since the density of
the JVPs also grows as $\sqrt{\Hab}$, this should result in
an average $\Delta H_m\propto\Hab$ as observed by the
global measurements. Another important question that our
findings raise is the stability of mesophase. On one hand the
liquid lamellas destroy the shear modulus and may act as a
source of dislocations that destabilize the solid
lamellas, reducing their $H_m$. On the other hand, a finite
solid-liquid surface tension \cite{Soibel-2000} may
stabilize the solid lattice against formation of liquid
lamellas.

In summary, we have shown that in contrast to its very weak
average influence, the crossing JV lattice has a major
effect on the \emph{local} melting process of the vortex
lattice in BSCCO. At low in-plane fields, the JV stacks reduce
the local melting field of the PV lattice by about 0.7~Oe.
As a result, within homogeneous regions of the sample,
melting is nucleated along the plains containing the JV
stacks forming a periodic lamellar solid-liquid or
solid-`gas' mesophase with periodicity and direction that
is controlled by the in-plane field. We suggest that the PV
liquid or `gas' lamellas can be as thin as a PV
`monolayer'. These findings raise a number of intriguing
theoretical and experimental questions including the
microscopic origin of the liquid nucleation along the JVs,
the properties of the 2D lamellas of vortex liquid or
`gas', and the possible destruction of the JV lattice in
the lamellar mesostate.

We thank A. E. Koshelev for fruitful discussions. This work was supported by the German-Israeli Foundation (GIF) and by the US-Israel Binational Science Foundation (BSF). EZ acknowledges the support of EU-FP7-ERC-AdG.

\bibliography{all2-abbrev}
\end{document}